\documentclass[10pt,conference]{IEEEtran}
\IEEEoverridecommandlockouts
\usepackage{cite}
\usepackage{amsmath,amssymb,amsfonts}
\usepackage{algorithmic}
\usepackage{graphicx}
\usepackage{textcomp}
\usepackage{xcolor}

\usepackage{comment}
\usepackage{url}

\usepackage{relsize,etoolbox}% http://ctan.org/pkg/{relsize,etoolbox}
\AtBeginEnvironment{quote}{\small}% Step font down one size relative to current font.

\def\BibTeX{{\rm B\kern-.05em{\sc i\kern-.025em b}\kern-.08em
    T\kern-.1667em\lower.7ex\hbox{E}\kern-.125emX}}

\newcommand{\aspas}[1]{{``#1''}}
\newcommand{\mcode}[1]{$\tt #1$}

\newcommand{\lessonone}{It is possible to have students at different levels of education working on the same projects since teachers separate their activities and adjust the dunning level}
\newcommand{\lessontwo}{We must encourage students to use their communications skills, but it is recommended to have a plan B just in case of relationship problems}
\newcommand{\lessonthree}{If we make it clear for all stakeholders, from the beginning, what the objectives of the project are, it is possible to reduce the frustration of not having the system deployed}
\newcommand{\lessonfour}{It is wise to take precautions on legal issues related to the license to use and own software that may be produced, to avoid false expectations}
\newcommand{\lessonfive}{The mediating teacher has to spend a lot of time managing expectations to avoid frustration}
\newcommand{\lessonsix}{Use structured interviews and clear ranking criteria to measure the engagement and select potential external stakeholders}
\newcommand{\lessonseven}{We must adapt our course considering known issues and institution deadlines, having in mind that unknown situations may also happen}
\newcommand{\lessoneight}{Communication issues are part of the real experience and it is worth letting students have a more proactive attitude to get feedback from external stakeholders}
\newcommand{\lessonnine}{Mediating and collaborating teachers should discuss technical issues related to the projects to align their speech}

\begin{document}

\title{Supporting Real Demands in Software Engineering with a Four Steps Project-Based Learning Approach}
%\title{Supporting Real Demands in Software Engineering with an Integrated Project-Based Learning Approach}
%\title{Applying Project-Based Learning in Software Engineering to Support Real Demands}

\author{\IEEEauthorblockN{1\textsuperscript{st} Leonardo Humberto Silva}
\IEEEauthorblockA{\textit{Information Systems (IFNMG)} \\
\textit{Federal Institute Northern Minas Gerais}\\
Salinas, Brazil \\
leonardo.silva@ifnmg.edu.br}
\and
\IEEEauthorblockN{2\textsuperscript{nd} Renata Xavier Castro}
\IEEEauthorblockA{\textit{Student Monitoring (IFNMG)} \\
\textit{Federal Institute Northern Minas Gerais}\\
Salinas, Brazil \\
renata.castro@ifnmg.edu.br}
\and
\IEEEauthorblockN{3\textsuperscript{rd} Marice Costa Guimaraes}
\IEEEauthorblockA{\textit{Budget and Finance (IFNMG)} \\
\textit{Federal Institute Northern Minas Gerais}\\
Salinas, Brazil \\
marice.costa@ifnmg.edu.br}
}

\maketitle

\begin{abstract}
Project-based learning (PBL) is a student-centered and learn-by-doing approach that organizes learning around projects. While entrepreneurship and PBL in SE education are thrilling research topics, there seems to be very little work focusing on the pros and cons of involving external stakeholders to support real demands in software engineering education. Working on real projects also supports students to acquire leadership skills, such as communication, project management, and teamwork. This paper describes a case study integrating students from different Software Engineering programs and involving external stakeholders, underpinned by PBL concepts. We present how this study was designed and implemented in a large institution, in four steps, summarized as follows: (I) requirements gathering and design; (II) information system development and implementation; (III) integration tests and deployment process; (IV) support and maintenance activities.

The study had the participation of 59 students from a professional technical course in step one, working in teams, and 10 undergraduate students from a Bachelor program in Information Systems in the following steps, working in pairs. Overall, the feedback from stakeholders and students exceeded expectations, although it increased the workload of teachers. We were able to distill a new set of lessons learned, and we expect that at least some of them will be useful for anyone implementing a similar course. As a consequence of this study, we plan to institutionally formalize the PBL course improvement process by defining specific outcomes and measurements.

\end{abstract}


\section{Introduction}
\label{sec:introduction}
Software Engineering (SE) deals with the application of systematic, disciplined, and quantifiable approaches to develop, operate, maintain, and evolve software~\cite{feathers2004, fox2013engineering}. Currently, SE subjects are present in several undergraduate courses, although there is no consensus on what methods should be used to teach them~\cite{se-mapping-2014, sbes2019-figueiredo}. Balancing theory and practice is a recurring challenge in Software Engineering Education. A key objective in a SE program is to provide students with the necessary tools to begin professional engineering practice~\cite{SE2014}. Additionally, SE students are expected to be able to choose and implement the development process best suited to the reality of a software development company or sector.

Projects are complex tasks, based on challenging questions or problems, that involve students in design, problem-solving, and 
decision-making activities. Project-based learning (PBL) is a student-centered and learn-by-doing approach that organizes learning around projects. PBL allows students to work relatively autonomously over extended periods, culminating in realistic products or presentations~\cite{blumenfeld2011, Erdogan2015, CSEET-2016-ShutoWKFYO}. In SE education, PBL is one of the main successful methods broadly used~\cite{DelgadoCSEET2017, RupakhetiCSEET2017, MarquesIEEETE2018, sbes2019-figueiredo}.

Working on real projects also support students to acquire entrepreneurial skills, such as communication, project management, and team-work. In this scenario, one of the challenges regarding the adoption of projects based on real-world problems is to find stakeholders with whom the students can cooperate and who are able and willing to invest necessary time and resources~\cite{toce2018-steghofer, seet2019-burden}. Moreover, while entrepreneurship and PBL in SE education are burgeoning research topics, there seems to be very little work focusing on the pros and cons of involving external stakeholders in software engineering education.

This paper describes a case study integrating students from different programs in software engineering education, underpinned by project-based learning.  In this study, we divide development activities among students from different programs, at different levels of SE education, avoiding some pitfalls highlighted in the literature when involving external stakeholders in academic projects. We present how this study was designed and implemented in a large institution, in four software engineering steps. The first step is mostly related to requirements gathering and design. Step two is dedicated to information system development and implementation. Step three encompass integration tests and deployment processes. And step four is planned for support and maintenance activities for the deployed systems. We investigate the evolution of the course over four semesters through the analysis of team projects, student reports, instructor notes, and structured feedback. Our experience allowed us to unveil a rich set of lessons learned, which are will help to refine the course and, hopefully, serve as guidance for those trying to include a similar approach in their curriculum.

The remainder of this paper is organized as follows. Section~\ref{sec:pbl} presents the main concepts of PBL. Section~\ref{sec:pbl_description} describes the SE course where the study was performed. Section~\ref{sec:case_study_design} describes the goals, method, and the study design for the execution of this study. Section~\ref{sec:implementation} presents how our PBL approach was implemented in a large institution. Section~\ref{sec:lessons} presents the discussion of the lessons learned. Section~\ref{sec:related} discusses related work. Finally, Section~\ref{sec:conclusion} concludes this paper and indicates future developments.

\section{Project-Based Learning}
\label{sec:pbl}

Project-based learning (PBL), also known as problem-based learning, is a learn-by-doing approach to science education that focuses on helping students develop self-directed learning skills~\cite{boud1998challenge, sahyar2017}. In SE education, PBL is one of the main successful methods broadly used~\cite{DelgadoCSEET2017, RupakhetiCSEET2017, MarquesIEEETE2018, sbes2019-figueiredo}. PBL transcends the classroom, involving students in the investigation of realistic problems, learning by working on projects, discovering and finding solutions as they progress along the path. Brender~\cite{bender2012project} describes PBL as an instructional model based on having students facing real-world issues and problems that they consider significant, determining how to solve them and then, acting collaboratively to create problem solutions. The instructor has a less central role, acting as a mediator, and students take more responsibility for their learning. Proposals for interdisciplinary activities are favored and encouraged in PBL, which also gives students the opportunity to work relatively autonomously over extended periods, culminating in realistic products or presentations~\cite{blumenfeld2011, Erdogan2015, CSEET-2016-ShutoWKFYO}. 

Despite the benefits of PBL, there are some difficulties and lessons learned in the related literature regarding the application of PBL and verification of its results~\cite{oliveiraSG13, jazayeri15, DelgadoCSEET2017, csedu2018, toce2018-steghofer, seet2019-burden}. Project-based learning does not have a fixed structure, making its implementation sometimes ad-hoc and opportunistic. Protocol definition and working procedures require more time for preparation and operation. Teachers have to manage stakeholders' expectations and frustrations. On the other hand, the learning process is
more flexible and involves interaction and cooperation among students, teachers, and other stakeholders. Moreover, the outcomes achieved at the end motivate learners and educators to create better projects in the future~\cite{pblweb2001, sahyar2017}.

\section{PBL Course Description}
\label{sec:pbl_description}

This PBL course relies on the cooperation of four different programs that encompass software engineering content, at a large public educational institution:  
\begin{itemize}
	\item A Professional and Technical Computer high school program (\aspas{PTC program}, henceforth)
	%, which offers an introduction to SE in its program, with involves the following subjects in this project: Analyses and Systems Design (APD), Systems Development (SD), and Database (DB).
	\item An Information Systems bachelor undergraduate program (\aspas{IS program}, henceforth)
	%, which offers the following subjects that are directly involved in this project: Software Engineering I (SE I), Software Engineering II (SE II), Object-Oriented Programming I (OOP I), Object-Oriented Programming II (OOP II), and Database I (DB I). 
	\item A Supervised Curricular Internship program (\aspas{SCI program}, henceforth)
	\item A Junior Enterprise program (\aspas{JE program}, henceforth)
\end{itemize}

\vspace{1.5mm}

PTC is a well-known established program that aims to train high school technical professionals to perform activities in the IT area, acting as ethical, critical, and apt citizens before entering higher education. The content of this program is not as comprehensive and in-depth as in an undergraduate education matrix, but includes the main concepts related to algorithms, object-oriented programming, database, and system analysis with software engineering concepts.    
%"Formar cidadãos-profissionais técnicos de nível médio para o exercício de atividades na área de informática, bem como, para atuar como cidadãos éticos, críticos e aptos para o ingresso no ensino superior. " 

For the IS program, the SE course has 160 hours divided in two subjects, Software Engineering I (SE-I) and Software Engineering II (SE-II), with 80 hours each. The course syllabus includes: software engineering concepts, software product requirements, life cycle, and software development paradigms, software quality, agile methods, detailed software design and design patterns, software architecture and structure, validation and verification, system implementation and tests, software maintenance and configuration management. The objectives of creating projects that involve students from PTC and IS programs are: (i) encourage teamwork by integrating students from the technical course with students from the higher education course; (ii) apply the same projects in both programs allowing all involved students to benefit from PBL learning; (iii) divide tasks to create realistic schedules allowing students in the IS program to have time to finish the software they produce. 
%The information systems undergraduate course was recently created in our institution, having the first class initiating in the first semester of 2017. 

The SCI program complements the process of teaching / learning through the application of scientific-technical knowledge in real situations during the exercise of a future profession. Students must fulfill the minimum of 200 hours of an internship to be eligible for graduation. The idea is to use the internship to deploy systems that were successfully completed at the end of the IS program. This step aims to tackle one of the frustrations mentioned by Stegh{\"{o}}fer et al.~\cite{toce2018-steghofer}  when university staff is involved in the PBL process, with great expectation, and sees the product been discarded when the project is finished. Among the goals of the SCI program, the following stand out: 

\begin{itemize}
	\item Enable and encourage students to increase professional training.
	\item Know the philosophy, guidelines, organization, and operation of companies and institutions.
	\item Improve interpersonal relationships and teamwork skills.
	\item Exercise critical sense and creativity in the future profession. 
	\item Participate in projects, research, and/or extension programs within the scope of professional practice. 
\end{itemize}

\vspace{1.5mm}

It is important to highlight that not all projects will be deployed, but just those that reach a quality level that is approved by all stakeholders. Besides, students can choose to do their internships in other areas and/or companies. In these cases, a system developed and approved by the stakeholders would have to wait for another team that could take over the deployment stage. 

Once students graduate they tend to move on, leaving the systems they deploy behind. To give continuous support to systems deployed by the SCI program, students from the JE program take control and keep supporting clients maintaining and evolving the system. Junior enterprises (JEs) are organizations managed by undergraduate students and designed to be training spaces for entrepreneurship and professionalization in connection with the external community. The primary goal of JEs is to develop their members personally and professionally through business experience, carrying out projects and services in the area of expertise of the undergraduate course to which the junior company is linked. They offer low-cost services in different areas of knowledge, and students' activities are guided by professors from the host institution. Thus, in addition to reaching their own goal, the JEs contribute to the development of entrepreneurship in their community. JEs mainly serve micro and small companies, which usually do not have access to senior consultancy when facing management difficulties. Moreover, participating in a junior enterprise, as part of extra-curricular activities, complement rather than substitute entrepreneurship education~\cite{almeida2019}.
%The quality of student work and its regular offer are positive aspects for local development and for the sake of entrepreneurial culture.

%Objetivos para integrar TC com IS:
%1 - possibilitar trabalho em conjunto integrando alunos do curso técnico com alunos do curso superior.
%2 - aplicar, ao mesmo tempo, os mesmo projetos PBL in both programs 
%3 - dividir tarefas para permitir que os alunos do programa SE tenham tempo viável para realmente terminem os softwares e
%poderem complementar sua formaçao participando de projetos de extensão das etapas de treinamento e implantaçnao de sistemas. 

\vspace{1.5mm}

\noindent The course initiated in 2018 and since the beginning we adopted PBL principles as follows:

\begin{itemize}
	\item \textit{Project Based}: There is a software development project that is central to the course. Most classroom activities and lectures are driven by the progression of the project.
	\item \textit{Realistic}: Each project represents a real-world problem, and we use institution employees acting as real customers (external stakeholders) during the course.
	\item \textit{Evidence-based}: The students have to deliver intermediate products, such as text notes, diagrams, and source code, mostly related to the software development life cycle. All activities of the project are driven by meaningful questions that direct students into investigating and applying the SE theory for the project.
	\item \textit{Teamwork}: The students have to work in teams, typically from 4 to 6 students in the PTC program e from 2 to 4 students in the IS program.
	
\end{itemize}

\section{Case Study Design}
\label{sec:case_study_design}

The PBL approach for this study is divided into four steps: (i) Requirements gathering, starting with PTC program; (ii) System development, moving to IS program; (iii) System deployment, implemented by students enrolled in the SCI program, and (iv) Maintenance and support activities, implemented by members of the JE program. The stakeholders for each project are distributed into the following groups:

\begin{itemize}
	\item Mediator: professor that coordinates the course.
	\item Collaborators: professors responsible for subjects involved in the programs. 
	\item Clients: University staff members that have specific demands.
	\item System Analysts: Students responsible for executing the project, under supervision. 
\end{itemize}

\vspace{1.5mm}

As initial planning, the mediator meets with other professors (Collaborators) to define how to conduct and evaluate the projects. Each professor must allocate part of their classes to explain the program and, during each semester, allow the development of activities related to the project. Students, especially from the PTC program, have very little or no free time at all. Therefore, the projects cannot be treated as extra activities and must be carried out during the period of classes. This is imperative for the course to be successful and not to interfere with other extracurricular activities.

After meeting collaborators, the mediator makes contact with university staff members to select potential clients and projects. This selection depends mostly on the sector's demands, which could benefit from the use of an information system. The client's availability is also important once they need to take part in project meetings and other events. The selected clients are then briefed about how the PBL approach works. Since this moment, it is important to emphasize that the goal is to allow the students to participate in a real situation of system development, making clear that the final product, software in this case, is not the main objective. 

Once projects and clients are defined, it is time to discuss the idea with the students. It is important to motivate them, making them understand that they are going to play the role of system analysts, discussing a real demand with real clients. Again, it has to be clear that they will be evaluated during monitoring activities as the projects evolve, independently if the final product will be deployed or not.

After the initial clarification phase, students from the PTC program are divided into teams (4 to 6 members) and each team must appoint one of them to be the project manager. As in the industry, the manager's leadership role is of great importance in motivating people and creating an effective working environment, and we intend to encourage that among students. After organizing the teams, it's time to allocate a project to each one. We established that each team, belonging to the same class, should have a distinct project. Different projects have different levels of difficulty and complexity, and this has to be considered when guiding and evaluating teams.

After project allocation, students will be able to work on them during classes, depending on the strategy defined by the collaborators. The theory learned in class will be used directly in the projects. Subjects related to software engineering and software development take part in this step. Once these subjects are completed, by the end of the semester, it is expected that the students deliver the artifacts presented in Table~\ref{tab:artifacts}.

\begin{table}[htbp]
	\caption{Artifacts to be produced by PTC program (step one)}
	\begin{center}
		\begin{tabular}{|c|l|}
			\hline
			%& &  \textbf{Sector}   \\
			\textbf{ID}  & \textbf{Artifacts description}  \\
			\hline
			A1 & Detailed description of the system.  \\
			A2 & Functional and non-functional requirements.  \\
			A3 & Use case diagrams (UCs), representing user's interactions.   \\
			A4 & Entity Relationship Diagram (ERD) and Relational Model (RM).   \\
			A5 & System prototype with main screens (layout only).  \\
			\hline
			%\multicolumn{3}{l}{$^{\mathrm{a}}$Sample of a Table footnote.}
		\end{tabular}
		\label{tab:artifacts}
	\end{center}
\end{table}

The development of the first three artifacts demands knowledge of key concepts present in the syllabus of Systems Analysis and Design subject. It requires the use of communication skills and the ability to abstract, synthesize, and document information received by the user (client), who is probably not an IT person. We consider this a new and profitable experience that the students would not have in a traditional SE course, in which the teacher plays the rule of a user. To serve as example for the students, we provide real lists of requirements and UML use case diagrams~\cite{uml2006}. Artifacts A4 are related to the content of the Database subject. Students are asked to produce entity relationship diagrams and relational models. Finally, to produce artifact A5 (prototype) they need concepts present in Systems Development subject. The deadlines of the projects need to be synchronized with the subjects involved, allowing students to immediately apply the theory they learn in their projects and allowing teachers to assess students in their subjects. Teams are also asked to write a text describing their work, what was accomplished, and what still needed to be done. This documentation is the starting point for the next teams of the IS program that will take over the projects. 

\begin{figure}[htbp]
	\centerline{\includegraphics[width=3.6in]{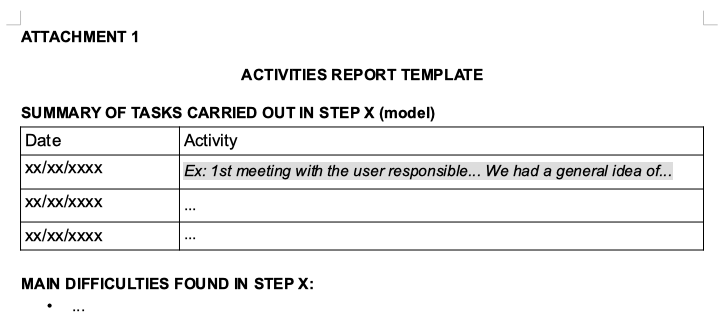}}
	\caption{Template recommended for weekly reports.}
	\label{fig:template}
\end{figure}

Along with the delivery of the mentioned artifacts, teams had to produce a weekly report resuming their activities, including meetings, main difficulties, achievements, and any other relevant information. Fig.~\ref{fig:template} presents a template we recommended to write these reports, which are very important for monitoring activities. This template consists of a text document, with a table to specify dates and activities carried out and / or scheduled. Besides, there is a space at the end to list the main difficulties encountered.

In step two, students from the IS program, guided by pair programming techniques~\cite{ppi2002, epe2004}, replace those from the PTC program in the System Analyst group. Similarly, professors in the Collaborators group will be replaced by those from the IS program. They all have to be briefed about the projects and the produced artifacts. It is expected that professors and students from both programs interact at this point. One can criticize the fact that students from the IS program do not participate in the initial discussions about the project, gathering requirements. However, we understand that if the students had to take part in all software development life cycle phases they probably would not be able to complete the software and reach the deployment step. Besides, the opportunity to analyze and understand artifacts produced by another team is a positive experience that highlights the importance of the quality of the artifacts, \emph{e.g.},~application of best practices in source code, concise and clear documentation, and detailed diagrams~\cite{benedetto-seet2020}. Besides, during the development of a system, it is normal to make changes and adaptations, meaning that students from the IS program are likely to update and improve the artifacts they receive. As happened in step one, students use the theory learned in class directly into their projects.
%, and subjects related to software engineering and software development take part in this step.

To go to step three, it is expected that students finish the project and present the outcome to the stakeholders. Projects eligible for deployment go to step three, in which the students involved enroll for the SCI program. The decision about which projects go to step three has to be made together with the agreement of all stakeholders. It is possible to have good projects that don't go to step three because the students want to do their internship with something else, and it is also possible to have projects that do not fulfill properly client's requirements. The outcome of step three is the deployment of the system in one specific sector of the institution. 

In step four, once the information system is successfully deployed, all artifacts and technology developed are transferred to the members of the JE program, including repository access. From this point, the junior enterprise takes over maintenance and support tasks.

\section{Case Study Implementation}
\label{sec:implementation}

This section describes how the study was implemented in a large institution, according to the methodology described in Section~\ref{sec:case_study_design}.

\subsection{PTC-Program Implementation (step one)}
\label{sec:ptc_implementation}

This implementation step was initiated in the second semester of 2018, involving the following subjects of the PTC program: Systems Analysis and Design, Systems Development, and Database. Two classes were involved, class one (\aspas{PTC-A}, henceforth) with 30 students, and class 2 (\aspas{PTC-B}, henceforth) with 29 students. After contacting university staff members, we selected seven proposals for projects. A brief description of the projects can be found in Table~\ref{tab:projects}.

\begin{table}[htbp]
	\caption{Selected Projects}
	\begin{center}
		\begin{tabular}{|c|l|l|}
			\hline
			%& &  \textbf{Sector}   \\
			\textbf{ID}  & \textbf{Brief description} &  \textbf{Sector}  \\
			\hline
			P1 & Information system for soil analysis & Soils Lab \\
			P2 & Production control and input distribution system  & Financier \\
			P3 & System for managing internships  & Internship \\
			P4 & Production control system for agribusiness  & Agribusiness \\
			P5 & Medical care control  & Nursing \\
			P6 & Appointments control for the veterinary hospital  & Veterinary H. \\
			\hline
			P7 & Research Projects Support System$^{\mathrm{a}}$  & Research \\
			\hline
			\multicolumn{3}{l}{$^{\mathrm{a}}$This project was included later in step one.}
		\end{tabular}
		\label{tab:projects}
	\end{center}
\end{table}

Classes were then divided into groups (teams) and each team was assigned to a project. The decision on how the teams would be composed was delegated to the students. Class PTC-A was divided into five teams with six students each. Projects P1, P2, P3, P5, and P6, from Table~\ref{tab:projects}, were assigned to the teams of PTC-A by a lot.  Class PTC-B was divided into six teams, one of them with four students and the others with five students each. Projects from P1 to P6 were assigned to teams of class PTC-B, also by lot. Table~\ref{tab:assignment} summarizes that assignment of projects to teams. Only project P4, in class PTC-A, was not assigned to any team.

\begin{table}[htbp]
	\caption{Initial Allocation of Projects to Teams}
	\begin{center}
		\begin{tabular}{|c|c|c|c|}
			\hline
			\textbf{Class}  & \textbf{Project} &  \textbf{Team} & \textbf{\# Students}  \\
			\hline
			& P1 & T1 & 6 \\
			\cline{2-4} 
			& P2 & T2 & 6 \\
			\cline{2-4} 
			& P3 & T3 & 6 \\
			\cline{2-4} 
			PTC-A & P4 & - & - \\
			\cline{2-4} 
			& P5 & T4 & 6 \\
			\cline{2-4} 
			& P6 & T5 & 6 \\
			\cline{2-4} 
			& \multicolumn{2}{|r|}{Total of Students} & 30 \\
			\hline
			& P1 & T6 & 5 \\
			\cline{2-4} 
			& P2 & T7 & 5 \\
			\cline{2-4} 
			& P3 & T8 & 4 \\
			\cline{2-4} 
			PTC-B & P4 & T9 & 5 \\
			\cline{2-4} 
			& P5 & T10 & 5 \\
			\cline{2-4} 
			& P6 & T11 & 5 \\
			\cline{2-4} 
			& \multicolumn{2}{|r|}{Total of Students} & 29 \\
			\hline
		\end{tabular}
		\label{tab:assignment}
	\end{center}
\end{table}

Projects were monitored during the development of the artifacts described in Table~\ref{tab:artifacts}. For each artifact, we describe next the main observations and feedback collected from classes, meetings, statements, weekly reports, and from the artifacts themselves.

\vspace{1.5 mm}

\noindent \textit{Artifacts A1 (system description) and A2 (requirements).} Most teams (9 out of 11) described in their weekly reports that the main difficulty related to these artifacts was: \aspas{\textit{To understand and document everything reported by the user}}. Team T1 requested to change their project, arguing that they did not like project P1 and were not able to understand the requirements. We could notice that the students were not getting along well with the user of project P3. And this same team (T1) also refused to work with project P4, that was available for class PTC-A. Because of this unexpected situation, the mediating teacher searched and introduced a new project (P7 - see Table~\ref{tab:projects}), which was assigned to the team T1, with no objections this time. 

\vspace{1.5 mm}

\noindent \textit{Artifact A3 (use cases).} During the development of the UML use cases, the main technical difficulty reported was related to the specification of alternate and exception flows in each diagram, as reported by team T6: 

\begin{quote} \aspas{The main difficulties we had were (1) elaborating alternative flows for normative requirements; (2) understanding exceptions in the use case specification.}\\ (Team Report)\end{quote} 
Most teams had similar reports. Moreover, three teams (T2, T6, and T8) reported relationship problems and disagreements among members. We had to reallocate two students from their original teams because of that. A student from the T6 team switched places with another from the T8. 
%\noindent \textit{Artifact A3 (use cases).} During the development of the use cases, the main technical difficulty reported was related to the specification of alternate and exception flows in each diagram, as reported by team T6: \aspas{\textit{The main difficulties were (1) elaborating alternative flows for normative requirements; (2) understanding exceptions in the use case specification}}. Most teams had similar reports. Moreover, three teams reported relationship problems among members. We had to reallocate two students from their original teams because of that. 

\vspace{1.5 mm}

\noindent \textit{Artifacts A4 (entity relationship and relational models).} These artifacts were developed while students learned how to use modeling tools such as  \mcode{Astah}\footnote{\url{https://astah.net/}} and \mcode{MySQL}  \mcode{Workbench}\footnote{\url{https://www.mysql.com/products/workbench/}}. At this stage, the main difficulties varied considerably, depending on the project. For example, projects P2 and P6 had more complex entity relationships. Project P5 demanded a small number of tables but some of them with a large number of fields, as reported by team T4: \aspas{\textit{We are facing some difficulties with the size of the documents/entities}}. In general, students also reported some difficulties to determine the type and size of the table fields, \emph{e.g.},~ when to use the type \mcode{char} instead of \mcode{varchar}  or a \mcode{blob}  instead of a \mcode{text}.
 
\vspace{1.5 mm}

\noindent \textit{Artifact A5 (system prototype).} To complete this artifact students had to exercise mainly their programming skills. One of the difficulties reported was related to task allocation for the development of interfaces. We can see how team T2 dealt with this problem in the following excerpt of their report: 

\begin{quote}\aspas{Our group is satisfied with the results achieved so far. In some tasks, we had greater participation of some members, and in other tasks of others. We left the less complex parts for students with greater difficulty in programming.}\\ (Team Report)\end{quote} 
Team T1, which requested to change the project in the beginning, and also presented some internal relationship problems, summarized their participation as follows: 

\begin{quote}\aspas{At the beginning of the project there were several difficulties, however, we managed to fulfill our goals and all members helped equally in the programming. We are happy with the final result.}\\(Team Report)\end{quote} 
In short, all teams were able to deliver their prototypes, some very good, others not so much, but the evaluation considered the evolution during the entire development process.

\subsection{IS-Program Implementation (step two)}
\label{sec:is_implementation}

In this second step, students from the IS program replace those from the PTC program inheriting their projects, and all artifacts produced so far. This step initiated in the first semester of 2019, involving the following subjects of the IS program: Software Engineering, Object-Oriented Programming II, and Database. One class was involved (\aspas{IS-A}, henceforth) with 10 students.

Class IS-A was then divided into groups (pairs) and each pair was assigned to a project. The decision on how the pairs would be composed was delegated to the students. There were five pairs in total. Students were then introduced to the projects, and its artifacts, so they could decide which ones they wanted to take over. Table~\ref{tab:is_assignment} presents the assignment of projects to pairs. The first pair (PR1) chose to continue project P2, started by team T2 from the PTC program. Pair PR2 chose project P5, started by team T4, while pair PR3 chose the same project, but with artifacts developed by team T10. Pair PR4 chose project P6, started by team T5, and pair PR6 also chose P6 but started by team T11.

\begin{table}[htbp]
	\caption{Initial Allocation of Projects to Pairs in the IS program}
	\begin{center}
		\begin{tabular}{|c|c|c|}
			\hline
			\textbf{Class}  & \textbf{Pair} &  \textbf{Project}  \\
			\hline
			& PR1 & P2 by T2 \\
			\cline{2-3} 
			& PR2 & P5 by T4 \\
			\cline{2-3} 
			IS-A & PR3 & P5 by T10 \\
			\cline{2-3} 
			& PR4 & P6 by T5 \\
			\cline{2-3} 
			& PR5 & P6 by T11 \\
			\cline{2-3} 
			\hline
		\end{tabular}
		\label{tab:is_assignment}
	\end{center}
\end{table}

The first two weeks were intended to allow pairs to know their projects in more detail. They were encouraged to interact with the user and also with the students from the PTC program, to better understand systems requirements and artifacts. During this period, PR2 requested to change their project, stating that the artifacts they chose initially are no so good and that the source code was too complicated. They decided to move to project P2 by team T7.

With exception of PR1, all others complained about the quality of source code and commentaries they received, stating that many things had to be rebuilt, as we can see in this excerpt from the first report of pair PR4: 

\begin{quote}\aspas{We will keep the information present in the documents as well as the text information on each screen of the program. However, we have to refactor classes into proper packages, as well as separate the images and put all interfaces and buttons to work.}\\(Pair Report)\end{quote} 
PR5 argued that refactoring was not worth and decided to rewrite the source code from scratch: 

\begin{quote}\aspas{After analyzing the code, it was verified that it is not feasible to modify the existing code, and it was therefore decided to create a new one. The layout and functionality for the time being, will be the same.}\\(Pair Report)\end{quote} 
Summarizing, students from the IS program were satisfied with documentation, diagrams, and requirements gathered in step one, but the artifacts related to source code and database tables were difficult for them to understand and maintain. Only PR1 praised the source code they inherited.

\subsection{SCI-Program Implementation (step three)}
\label{sec:sci_implementation}

In this third step, stakeholders of each project get together to evaluate and select which ones are eligible for deployment. This step was initiated in the second semester of 2019. From the five projects of step two (IS program), three of them were selected: P2 (pair PR1), P2 (pair PR2), and P5 (pair PR3). As presented in Table~\ref{tab:projects}, project P2 was designed for the Financial department of the institution, and, at this point, there were two solutions available for this sector. We had at least three meetings to decide which solution would be deployed. The employees involved could not make the decision, and it was necessary to establish specific criteria in order to choose. In the end, we decided to continue with P2 (PR1) and P5 (PR3).  

To deploy P2, students initially faced a problem with the computer desktop available in the financial department. When the system was executed, the screens always appeared truncated, cut, and it was necessary to adapt the layout to support older operating systems and monitors. In addition to this problem, there was a period in which the sector was involved in a deadline and couldn't support the students in the internship and, because of that, the deployment was delayed. 

The deployment of P5 inside the infirmary of the institution demanded a series of changes in the database tables, given the issues that arose when users started entering data. It was necessary to go back a few steps in the specification to adapt to the system. The participation and engagement of the responsible nurse were essential for the final validation of the system. To install the database in one of the institution's MySQL servers, the students also had to interact with employees in the information technology department. This was their first experience of deploying an internal information system developed by students in a teaching project. All protocols were new, for teachers, students, and employee members of IT department.

\subsection{JE-Program Implementation (step four)}
\label{sec:je_implementation}

When classroom lessons were interrupted, due to the pandemic of {COVID-19}, we had to stop our PBL course before creating the junior enterprise that would receive, store, maintain, and support the systems deployed. The institution has a specific regulation for the creation of companies linked to undergraduate courses, and all preparations were already being made to formalize and complete this step. The intention is to resume this initiative, where it left off as soon as face-to-face activities can return safely.

\section{Lessons Learned}
\label{sec:lessons}

In this section, we provide an overview of the lessons we have learned from applying our PBL approach. We describe these lessons according to the  seven recurring risk themes described in~\cite{toce2018-steghofer}  when involving external stakeholders in project courses: student ability, outcome, expectation, engagement, context, feedback, and misalignment.

\vspace{1.5 mm}

\noindent \textit{Student Ability.}  This risk is associated with the knowledge, skills, and abilities of the students. In our approach, we worked with two different groups of students: we started with high school level students from a professional technical course in the PTC program and then continued with undergraduate students for the rest of the development process. The students of the technical course do not see the contents of the subjects with the same depth as the undergraduate students. For example, subjects involved in the PTC program do not encompass Scrum~\cite{scrum2012} and Kanban~\cite{kanban2015} agile fundamentals, which are present in the undergraduate syllabus. For this reason, the dunning level must be different, and teachers were able to evaluate them separately since these different groups of students do not work together. Students in the PTC program start in step one, and after that, the projects continue with undergraduate students in the following steps. The output of one step is the input for the next one, and the different teams do not have to interact, avoiding problems that could arise from the heterogeneity of the students involved. Overall, all groups of students could benefit from the learning process. Some of them who showed no interest in the theoretical classes, before the PBL course, took on a leadership role in the project activities and automatically improved their performance.

\vspace{0.5 mm}
\noindent{\em Lesson \#1: \lessonone.}
\vspace{1.5 mm}

In addition to the technical abilities required of students, we understand that communication issues play an important role here. The effective participation of users is fundamental to the evolution of the projects, and communication skills are also required of students, which represents a major challenge for most of them, in all steps of the project implementation. Students had important lessons about how to conduct meetings and gather relevant information. Besides, unexpected problems in this field may occur, as we presented in Section~\ref{sec:ptc_implementation}, when teachers had to intervene and reallocate students and tasks due to internal and external relationships and communication problems. It is important to show students that a good relationship with team members and external stakeholders is key to success. Moreover, a very recent study with Stack Overflow jobs demonstrated that communication, collaboration, and problem-solving are the most demanded soft skills that IT companies look for in candidates~\cite{ist2020}. 

\vspace{0.5 mm}
\noindent{\em Lesson \#2: \lessontwo.}
\vspace{1.5 mm}

\noindent \textit{Outcome.} This risk is associated with the desired results for all lenses: students, external stakeholders, and teachers. At first, students have a certain fear of not being able to reach the end of the project with a satisfactory outcome, as it is a real demand, with real users.  External stakeholders, potential users, in this case, are not concerned about grades or been evaluated, they create instead an expectation related to the possibility of having an information system that will help them with their daily activities. For the teachers, the priority is on student learning and completion of the related SE subjects. 

%If the process yields something that is useful for the institution, that is a bonus, but it is not the goal of the PBL approach. 

%As the project develops, we see a positive change in their behavior. Students get involved with the demands of the sector and a desire to deliver a product that is useful grows.

As highlighted before, the expectation around the outcome of a project is twofold, it feeds and increases stakeholders motivations, but, in the same way, it can create a feeling of disappointment if the system is not deployed at the end. We were able to experience this feeling, since of the seven sectors/projects involved, only two went through the deployment step. However, how the activities and meetings with external stakeholders were conducted, making it clear, from the beginning, what the main objectives were, contributed to the relief of frustration, mitigating this threat. It was possible to evidence this fact when projects ended, and people responsible for sectors that were not contemplated with the deployment of a system stated that they are willing to participate again in the next editions of the PBL implementation.

\vspace{0.5 mm}
\noindent{\em Lesson \#3: \lessonthree.}
\vspace{1.5 mm}

%Another point to be considered was the feeling of ownership that students demonstrated regarding their systems/projects. We often hear from them: \aspas{\textit{My system...}}, \aspas{\textit{My code...}}, \aspas{\textit{My product...}}, suggesting that the outcome of projects would be their property. During one of the monitoring activities, one student came with the following question: \aspas{\textit{If the system looks good, can we sell it afterward?}}. This question raised some concerns, on the behalf of teachers, about the legal implications of using a software developed by students in a SE course. We then realized that we should have spent more time discussing institutional regulations, ownership, software license, and terms of use with students and external stakeholders.

Another point to be considered was the feeling of ownership that students demonstrated regarding their systems/projects. We often hear from them: \aspas{\textit{My system...}}, \aspas{\textit{My code...}}, \aspas{\textit{My product...}}, suggesting that the outcome of projects would be their property. During one of the monitoring activities, one student came with the following question: \aspas{\textit{If the system looks good, can we sell it afterward?}}. This question raised some concerns, on the behalf of teachers, about the legal implications of using a software developed by students. We then realized that we should have spent more time discussing institutional regulations, ownership, software license, and terms of use with students and external stakeholders.

\vspace{0.5 mm}
\noindent{\em Lesson \#4: \lessonfour.}
\vspace{1.5 mm}

\noindent \textit{Expectation.} This risk is associated with the aims and motivations for: students, external stakeholders, and teachers. Initially, students are more concerned about academic achievements and having good grades. External stakeholders are willing to help with the educational approach, but they also expect to benefit from this interaction to enhance their daily activities and processes. Teachers expect students to learn from a fruitful experience of developing an information system and if this process yields something useful, even better. 

As the projects evolved, we observed that students got involved with the demands of the sectors, and a desire to deliver a product that was useful grew. Playing the role of clients, external stakeholders presented some unrealistic demands which were difficult for the students to understand. And some demands, although realistic, needed filtering to be developed in time. Furthermore, it is difficult to mediate the interaction between students and external stakeholders ensuring that these interactions are constructively aligned with the course objectives~\cite{biggs96}. To tackle these problems, the mediating teacher had to act and often propose design changes to make projects viable within the deadline and the purpose of learning through projects. In the end, we could achieve satisfactory results in managing expectations, but it took much more time and effort from the mediator than expected. 

\vspace{0.5 mm}
\noindent{\em Lesson \#5: \lessonfive.}
\vspace{1.5 mm}

\noindent \textit{Engagement.} This risk is associated with the effort, time, or resources the external stakeholder can invest in the course. In our study, we initially took some time to select potential projects to participate in. We interviewed external stakeholders involved to measure their availability and level of engagement. Nevertheless, we noticed that some dedicated themselves more than others during the activities, for many different and unforeseen reasons. In project P6, for example, the external stakeholder responsible told the students that he would be able to attend meetings and other monitoring activities only on Tuesday mornings. This clearly interfered with the progress of P6 activities. But this time constraint was established only after the beginning of the course. It was not detected during project selection interviews. This was an indication that our interviews should be better structured and guided by clearer ranking criteria.

\vspace{0.5 mm}
\noindent{\em Lesson \#6: \lessonsix.}
\vspace{1.5 mm}

\noindent \textit{Context.} This risk is associated with the institution or study program related issues that are beyond the teacher’s control. To mitigate this threat, we adapted our implementation to institution deadlines, calendars, pedagogical course design and regulations, and other known events scheduled to happen. Some things though are hard to predict, \emph{e.g.},~when we had two project outcomes, from different groups, approved for deployment in the same sector (Financial department - Project P2), and it was necessary to elaborate a new set of criteria in order to choose, as we presented in Section~\ref{sec:sci_implementation}. Another unexpected situation was the step four suspension due to the {COVID-19} pandemic, as we described in Section~\ref{sec:je_implementation}. Who could see that coming? 

\vspace{0.5 mm}
\noindent{\em Lesson \#7: \lessonseven.}
\vspace{1.5 mm}

\noindent \textit{Feedback.} This risk is associated with the interaction between external stakeholders and students, which is
crucial for the success of the course but hard to observe and control. The mediating teacher must monitor the students' activities, although it is necessary to give them some independence during activities that require communication with those who are in the role of customers/users. In Section~\ref{sec:ptc_implementation} we showed that team T1 requested to change their project, and we had to create a new project (P7) for them. That was due to communication issues between team T1 and the user responsible for project P1. The teacher could act quickly in this case, but more disagreements may likely have occurred in other teams and projects without being reported by students or external stakeholders. We invested in scheduling regular meetings and workshops as monitoring activities to detect relationship problems and provide valuable feedback to students. We believe that the communication issues involved are part of the real experience we wanted to provide to students because they will probably go through similar situations when they leave the academy and go to industry. We could also observe some students who previously did not show interest starting to have a leadership position in the project. Overall, the interaction between students and external stakeholders provided valuable feedback and contributed considerably to projects.

\vspace{0.5 mm}
\noindent{\em Lesson \#8: \lessoneight.}
\vspace{1.5 mm}

\noindent \textit{Misalignment.} This risk is associated with possible misalignment considering the goals of the different parties and the course. While the expectations of the external stakeholders, students, and of the teachers on their own can be perfectly reasonable, compared to each other they
can be misaligned. Students often expected clear-cut answers and solutions, while external stakeholders may have unrealistic demands and nuanced descriptions. To address potential misalignment of expectations we invested in clear communication even before projects start.

One particular situation was observed during the development of projects. Sometimes students asked a practical question to a collaborating teacher, inside the classroom, and then asked the same question to the teacher mediator, getting different answers and solutions. If we consider for example the database, there are many ways we can arrange tables and constraints among tables to solve the same problem. Students got confused when they got different solutions from teachers. And then we realized that, in our approach, we should have scheduled some moments in which mediating and collaborating teachers could discuss technical issues related to the projects, without sharing them with the students or external stakeholders because it probably would influence their decisions. The benefit would be only to align teachers' speech.  

\vspace{0.5 mm}
\noindent{\em Lesson \#9: \lessonnine.}
\vspace{1.5 mm}

\section{Related Work}
\label{sec:related}

Existing projects have evaluated the application of PBL methods to promote SE education and focused on different methods~\cite{blumenfeld2011, Erdogan2015, RupakhetiCSEET2017, MarquesIEEETE2018, binder-seet2020}. However, to our knowledge, none of them divide activities among students from different programs, at different levels of SE education, and have a consistent plan to avoid pitfalls highlighted in the literature when involving external stakeholders in academic projects. 

Benedetto and Navon~\cite{benedetto-seet2020} present an approach to exploiting team shuffling dynamics in a formal software design course to convey the importance of SE concepts, making students aware of the practical value of design-related activities in a system development process. They conducted an empirical study in which they analyzed students performance during software design classes. Evidence collected by the qualitative analysis of predefined one-on-one interviews showed that students had a better understanding of the concepts taught with the team shuffling approach than before. 

Our work also benefits from this idea of team shuffling when undergraduate students receive the artifacts produced in step one by other teams, and have to take over ongoing projects. 

Souza et al.~\cite{sbes2019-figueiredo} perform an opinion survey study to evaluate the students’ perception of the adoption of PBL in SE education. They use questionnaires to collect responses of 49 undergraduate students divided into two samples: 32 enrolled in an introductory SE course that uses PBL, and 17 students enrolled in a SE course that adopts a traditional teacher-centered learning method (non-PBL). The PBL principles adopted include projects based on real-world problems, in which the instructors play the role of customers during the course. Among the results, in general, students agree that it is important to use practical software development projects in the context of SE education, and that two of the most recurring negative aspects pointed by the non-PBL sample was the lack of orientation in activities and the lack of classroom activities to support the development of projects.

Burden et al.~\cite{seet2019-burden} report a case study on how the integration of entrepreneurial experiences into a software
engineering project course can benefit students to increase their team-work skills and competencies. They discuss how to implement entrepreneurial experiences that focus on taking action and managing resources in agile software projects, making it possible to other SE educators to understand and even adopt specific course design aspects. Regarding SE education, one of the biggest challenges they encountered lies in finding appropriate stakeholders with whom the students can collaborate and who are able and willing to invest necessary resources.

We tackle the challenges they describe when we use university staff members to play the role of clients, and our study showed that it is possible to avoid known problems related to the involvement of external stakeholders in academic projects.

Delgado and Aponte~\cite{DelgadoCSEET2017} analyze the evolution of a PBL course in SE for undergraduate students. They monitor software project repositories and the students' feedback during a period of six semesters, to investigate how the adoption of a PBL method affects students’ grades and their project activities. Among the benefits, they conclude that the monitored students improved their grades and were able to add more functionalities in their applications. On the other hand, once the students had more time for improving the features in their applications, at the expense of close monitoring of internal code quality, it resulted in an increase in technical debt.

Stegh{\"{o}}fer et al.~\cite{toce2018-steghofer} develop a model that allows analyzing the involvement of external stakeholders in
university courses. They obtain insights from past course instances and use them to identify potential risks and benefits in future projects. They apply the model in eight courses in SE programs, each program using a different strategy to select the stakeholders. One of these strategies is the use of university employees acting as customers. After applying the model, they show that the students tend to take the projects more seriously when external stakeholders are involved, and that guest lectures and supervision can also decrease the workload for the teachers. However, there are challenges that affect students negatively. The authors observe that: (i) students are frustrated when stakeholders use different terminology than teachers; (ii) stakeholders are frustrated when students focus more on academic achievements than on fulfilling their needs; (iii) teachers are frustrated when students do not take the insights expected from the experience. Specifically in the case when university employees are used, they highlight that a plan for software maintenance is needed to enable its use. Therefore, the university’s IT staff must be involved before preparing the course project description. And users from different areas might have unrealistic requirements that can confuse the students.

%This related work was very important in our study because we used their findings to avoid pitfalls related to the involvement of external stakeholders.

\section{Final Remarks}
\label{sec:conclusion}

This paper described a case study integrating students from different SE programs and involving external stakeholders, underpinned by project-based learning. In this study, we divided activities of a software development process among the students, at different levels of SE education, and used a consistent plan to avoid some pitfalls highlighted in the literature when involving external stakeholders in academic projects. We presented how this study was designed and implemented in a large institution, in four steps, summarized as follows: (i) step one started in the second semester of 2018 and it was mostly dedicated to requirements gathering and design; (ii) step two initiated in the first semester of 2019 and it was dedicated to information system development and implementation; (iii) step three started in the second semester of 2019 and it was intended for integration tests and deployment processes; (iv) and step four, which was suspended due to the coronavirus pandemic, encompass support and maintenance activities for the deployed systems. The case study implementation had the participation of 59 students from a professional technical course in step one, working in teams, and 10 undergraduate students from a Bachelor program in Information Systems in the following steps, guided by pair programming techniques. Although we use different programs with students having different levels and learning objectives, each program takes part separately in one of the four steps described in our approach, avoiding problems that could arise from the heterogeneity of the students involved.
%which are run sequentially but independently.

We followed the idea of team shuffling when undergraduate students received the outcomes produced in step one by other teams. It is important to highlight some benefits of this approach: (i) since the end of project activities, in each step, has to be aligned with the end of the subjects involved in the program, if the students had to take part in all development tasks they probably would not be able to complete the software and reach the deployment step; (ii) the opportunity to analyze and understand artifacts produced by others stands out the importance of the quality of the artifacts, \emph{e.g.},~application of best practices in source code, concise and clear documentation, and detailed diagrams. In addition, the need to communicate with external stakeholders is present in all steps, and not just during the requirements gathering.

The involvement of external stakeholders in project courses was considered positive and produced very good results, but it came with some challenges, such as misalignment between stakeholders and students or difficulties to manage expectations. We designed and implemented our PBL approach to overcome these known challenges and provide a beneficial experience for the students in SE education. Overall, the feedback from stakeholders and students exceeded expectations, although it increased the workload of teachers. We were able to distill a set of lessons learned, some of which are echoed in related literature. We expect that at least some of them will be useful for anyone implementing a similar course. 

As a consequence of this study, we plan to institutionally formalize the PBL course improvement process by defining specific outcomes and measurements. Moreover, the creation of the junior enterprise for the Bachelor program in Information Systems suspended due to the pandemic crisis, in step four, will be used to meet other external community demands for software development, in addition to maintenance and support routines for the institution's internal systems.

%We highlight the lessons learned as follows: (i) \lessonone; (ii) \lessontwo; (iii) \lessonthree; (iv) \lessonfour; (v) \lessonfive; (vi) \lessonsix; (vii) \lessonseven; (viii) \lessoneight; and (ix) \lessonnine.

%\section{References}

\bibliographystyle{IEEEtran}
\bibliography{IEEEabrv,jseet2020}

\begin{comment}

\vspace{12pt}
\color{red}
IEEE conference templates contain guidance text for composing and formatting conference papers. Please ensure that all template text is removed from your conference paper prior to submission to the conference. Failure to remove the template text from your paper may result in your paper not being published.

\end{comment}

\end{document}